\documentclass[apjl]{emulateapj}
\usepackage{natbib}
\usepackage{graphicx}

\newcommand{\reduceme}{\mbox{R\raisebox{-0.35ex}{E}D%
\hspace{-0.05em}\raisebox{0.85ex}{uc}\hspace{-0.90em}%
\raisebox{-.35ex}{{m}}\hspace{0.05em}E}}

\newcommand{\z}{\ensuremath{{z}}}
\newcommand{\Reff}{\ensuremath{{R_\mathrm{e}}}}
\newcommand{\Msun}{\ensuremath{M_\odot}}
\newcommand{\Mstar}{\ensuremath{M_\star}}
\newcommand{\kms}{\ensuremath{\mathrm{km\,s}^{-1}}}

\newcommand{\Mdyn}{\ensuremath{M_\mathrm{dyn}}}

\begin{document}

\slugcomment{Accepted for publication in ApJ Letters}

\title{Velocity Dispersions and Stellar Populations \\
    of the Most Compact and Massive Early-Type Galaxies at Redshift $\sim$1}

\author{Jesus Martinez-Manso\altaffilmark{1}, Rafael Guzman\altaffilmark{1}, Guillermo Barro\altaffilmark{2}, Javier Cenarro\altaffilmark{3}, Pablo Perez-Gonzalez\altaffilmark{2,4}, Patricia Sanchez-Blazquez\altaffilmark{5}, Ignacio Trujillo\altaffilmark{6}, Marc Balcells\altaffilmark{6,7,8}, Nicolas Cardiel\altaffilmark{2}, Jesus Gallego\altaffilmark{2}, Angela Hempel\altaffilmark{6}, Mercedes Prieto\altaffilmark{6}}

\altaffiltext{1}{Department of Astronomy, University of Florida, Gainesville, FL 32611-2055; martinez@astro.ufl.edu}
\altaffiltext{2}{Departamento de Astrof\'{\i}sica, Facultad de CC. F\'{\i}sicas, Universidad Complutense de Madrid, E-28040 Madrid, Spain}
\altaffiltext{3}{Centro de Estudios de F\'isica del Cosmos de Arag\'on, Plaza San Juan 1, Planta 2, 44001 Teruel, Spain}
\altaffiltext{4}{Associate Astronomer at Steward Observatory, The University of Arizona}
\altaffiltext{5}{Universidad Aut\'onoma de Madrid, Ciudad Universitaria de Cantoblanco, 28049 Madrid, Spain}
\altaffiltext{6}{Instituto de Astrofisica de Canarias, 38200 La Laguna, Tenerife, Spain}
\altaffiltext{7}{Isaac Newton Group of Telescopes, Aptdo. 321, 38700 Santa Cruz de La Palma, Spain}
\altaffiltext{8}{Departamento de Astrof\'\i sica, Universidad de La Laguna, 38206 La Laguna, Tenerife, Spain}

\begin{abstract}
We present Gran-Telescopio-Canarias/OSIRIS optical spectra of 4 of the most compact and massive early-type galaxies in the Groth Strip Survey at redshift $z\sim1$, with effective radii $\Reff=0.5 - 2.4$ kpc and photometric stellar masses $\Mstar=1.2 - 4\times10^{11} \Msun$. We find these galaxies have velocity dispersions $\sigma = 156 - 236$ \kms. The spectra are well fitted by single stellar population models with approximately 1 Gyr of age and solar metallicity. We find that: i) the dynamical masses of these galaxies are systematically smaller by a factor of $\sim$6 than the published stellar masses using $BRIJK$ photometry; ii) when estimating stellar masses as 0.7$\times$\Mdyn, a combination of passive luminosity fading with mass/size growth due to minor mergers can plausibly evolve our objects to match the properties of the local population of early-type galaxies.
\end{abstract}


\section{INTRODUCTION}

The evolution of the structural properties of massive early-type galaxies (ETGs) is currently a topic of debate with many open questions. In recent years, studies have shown that galaxies with $\Mstar \geq 10^{11}\Msun$ are smaller by a factor of four at $\z \geq1.5$ \citep{daddi05,trujillo06,longhetti07,buitrago08,damjanov09} and a factor of two at $\z \sim1$ \citep{trujillo07, trujillo11} than the nearby population of the same mass. For the most compact and massive galaxies in particular, \citet{trujillo09} found that only $\sim$0.03\% of SDSS galaxies with $\Mstar \geq8\times10^{10} \Msun$ at $\z \leq0.2$ are ETGs with $\Reff \leq1.5$ kpc. This contrasts with the much higher fraction $\sim$15\% of such galaxies found at \z$\sim$1 by Trujillo et al. (2007, hereafter T07). Most interestingly, the luminosity-weighted ages of the local compact ETGs from \citet{trujillo09} are very low ($\sim$2 Gyr), which argues against this sample being the counterpart of a passively evolved population of high redshift galaxies. Thus, this suggests that most high redshift compact and massive galaxies must have undergone a systematic growth in size.\\ Among the different mechanisms proposed to explain this expansion, the one that reproduces best the observed evolution of the average mass-size relation \citep{trujillo11} is the minor merger scenario \citep{naab09,hopkins10}. With this mechanism, galaxies grow inside-out by accretion of smaller satellites that build up an extended envelope \citep{dokkum10,hopkins09}. This model predicts a substantial growth in mass and effective radius, with a mild decrease in stellar velocity dispersion $\sigma$. Thus, measurements of $\sigma$ at different redshifts are a direct way to constrain this evolutionary scenario. In addition, they allow for an independent mass estimate through the virial theorem.\\
The minor merger scenario is consistent with the common findings of $\sigma \lesssim 250$ \kms for general ETG samples at $\z=1-2$ \citep{gebhardt03,treu05,wel05,serego05,cenarro09,cappellari09,fernandezlorenzo11}. Conversely, it is greatly challenged by the dramatic evolution required for the super-dense galaxies ($\Reff<1$ kpc, $\Mstar \geq10^{11}\Msun$), which are expected to have much larger $\sigma$. \citet{dokkum09} measured  $\sigma \simeq 500\pm100$ \kms from a super-dense ETG at $z=2.2$, yielding a dynamical mass that is in agreement with the photometric mass estimate. However, more $\sigma$ measurements of these extreme galaxies are needed to support or weaken the need of a stronger evolutionary scenario. In this paper we present $\sigma$'s of four massive and compact ETGs at $\z \sim1$, two of which are super-dense. We compare dynamical/stellar masses and estimate the evolutionary paths of these objects. Throughout this work we adopt a standard cosmology of $\Omega_m=0.3$, $\Omega_\Lambda=0.7$ and $H_0=70$ \kms Mpc$^{-1}$.

\begin{figure}[h]
\includegraphics[scale=.62]{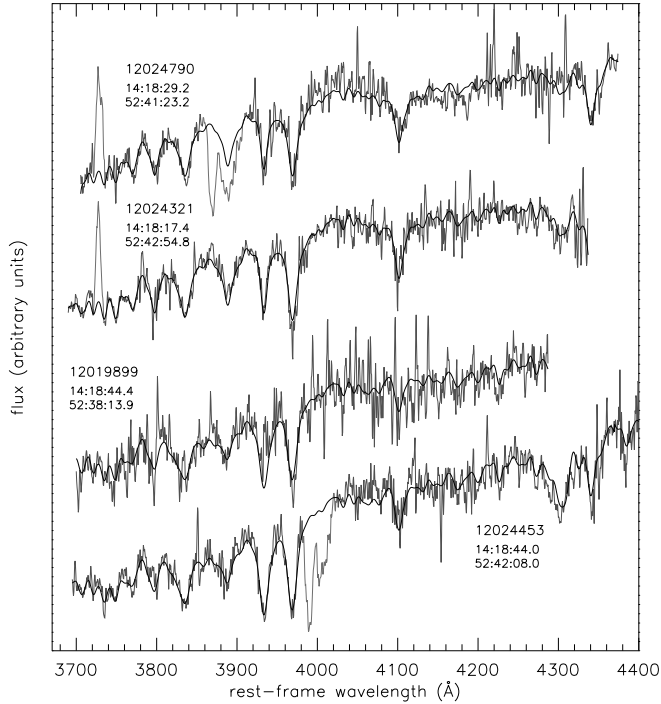}
\caption{\label{fig1}Spectra of our sample galaxies. The black solid lines represent the MOVEL fitting results. Light grey regions are [OII] emission at $3727\text{ \AA}$ and telluric absorption at $7600\text{ \AA}$, which were excluded from the fits (see text in \S 2.2). }
\end{figure}

\begin{figure}[h]
\includegraphics[scale=.77]{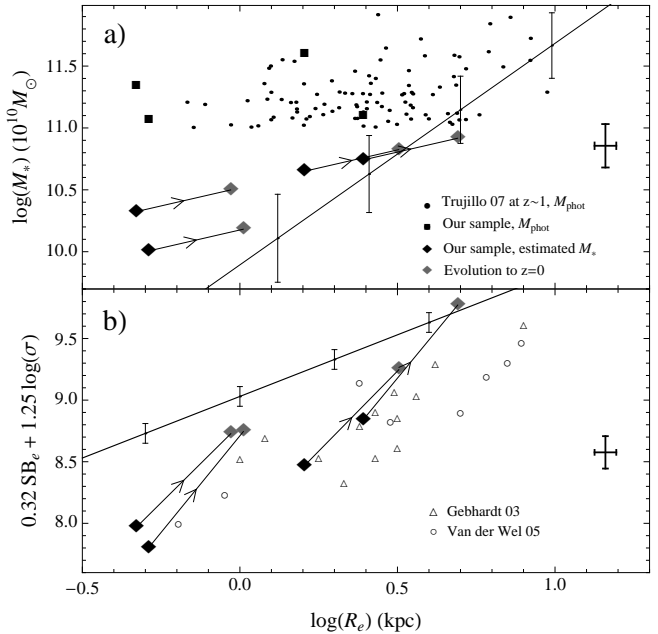}
\caption{\label{fig2}\textbf{a)} Stellar mass-size relation at $z\sim1$. Small circles represent the entire sample from T07, black squares are our selection among them. The straight line is the local relation from \citet{shen03}. Black diamonds are our galaxies with stellar masses derived as $0.7\,M_{\text{dyn}}$. Gray diamonds are their position when evolved to $z=0$ via minor merger growth. \textbf{b)} B-band Fundamental Plane of ETGs. The solid line is the local relation from \citet{jorgensen96}. Our sample is shown as black diamonds. Gray diamonds are their position when evolved to $z=0$ via minor merger growth and passive luminosity fading. Samples from \citet{gebhardt03} and \citet{wel05} in the range $z=0.8-1.1$ are shown for comparison. We transformed our $I$ to $B$ magnitudes appling \ensuremath{k}-corrections based on $V-I$ color as in \citet{gebhardt03}. In addition, we applied a negative offset of 0.1 mag on all surface brightness data from \citet{gebhardt03} to transform their Vega to our AB magnitude system.  }
\end{figure}

\section{THE DATA}

\subsection{Sample Selection}
Our targets were selected from the sample by T07, which contains photometrically derived parameters for 831 galaxies with $M_\star \geq10^{11}M_\odot$ up to $z\sim2$ in the EGS field from the POWIR/DEEP-2 survey \citep{bundy06}. Those authors measured effective radii with the GALFIT code \citep{peng02} and derived total stellar masses by fitting spectral energy distributions from \citet{bc03} with Chabrier IMF on $BRIJK$ photometry. We selected four massive and compact ETGs at $z\sim1$ with S\'ersic index $n>4$. Three of them have extreme densities in the sense that they are the most compact at a given mass. The catalog parameters of our selection are displayed in the first five columns of Table 1.

\subsection{Observations and reduction}

We observed 2 galaxies at a time using optical long-slit spectroscopy with the OSIRIS camera-spectrograph \citep{cepa98} at the 10.4 m Gran Telescopio Canarias. We used a 0.8$^{\prime \prime}$ slit and R1000 grism in the spectral range of $5000-10000$ \AA\ under a seeing of $\sim$0.7$^{\prime \prime}$. The total integration time per target was 12800 s. The instrumental resolution was determined by Gaussian fitting of the strong sky emission lines yielding $\sigma_\mathrm{inst}=120$ \kms. However, in one of the observing nights the spectrograph suffered a defocus that decreased the resolution, affecting half of the exposures of 12019899 and 12024453. For these frames we found $\sigma_{\mathrm{inst}}=210$ \kms, and we reduced this data subset independently. The standard star L970-30 was also observed to flux calibrate the spectra. The data reduction was performed with the IRAF longslit package and involved the standard steps of bias level correction, spectral flat-fielding, cosmic ray removal, sky subtraction, frame co-adding and aperture extraction. The telluric absorption band around 7600 \AA\ was present in all of our spectra, affecting the Ca H+K lines in targets 12019899 and 12024321. To correct for telluric absorption in these two galaxies, we used a F-type star observed in the same slit as 12019899. This star's spectrum has a signal-to-noise ratio $\mathrm{(S/N)}=280$ and a smooth continuum with no spectral features within the telluric band. After its continuum removal and normalization we obtained the atmospheric transmission profile in the spectral window and divided the galaxy spectra with affected Ca H+K lines by this transmission profile. We did not apply the telluric band correction on the other targets because the main spectral lines used to measure $\sigma$ were not affected in them. Note that this correction increases the noise in the spectral window where it is applied. Thus, in these cases the correction did not contribute towards more precise $\sigma$ measurements and for this reason we simply excluded the affected range in wavelength from the analysis. The final rest-frame spectra have $\mathrm{(S/N)}=22-29$ per \AA\ and are shown in Fig. \ref{fig1}. Note the strong absorption in the Ca H+K lines and the Balmer series from H$_\gamma$ to H$_{12}$. [OII]$\lambda 3727$ \AA\ doublet emission was found in two objects. In 12024790, this emission shows a double peak with a separation of 4.1 \AA\, as compared to 2.7 \AA\ of the doublet.

\begin{deluxetable*}{ccccccccccccc}
\tabletypesize{\scriptsize}
\tablecaption{Physical parameters of the observed targets\label{table1}}
\tablewidth{0pt}
\tablehead{
\colhead{ID} & Redshift & \colhead{$I$(ABmag)}& \colhead{$\Reff$(kpc)}&\colhead{log$M_\star(M_\odot)$} & \colhead{SB$_e$(mag)}  & \colhead{log$M^{R}_\star(M_\odot)$} &  \colhead{$\sigma$(km s$^{-1}$)}& \colhead{log$M_{\text{dyn}}(M_\odot)$} &\colhead{Age (Gyr)} &\colhead{[Fe/H]} }
\startdata
12024790 & 0.9656 & 21.89 & 0.514 & 11.07 & 15.82 & 10.65$\pm$0.03 & 156$\pm$10 & 10.16$\pm$0.08 & 0.8 & 0.0\\
12024321 & 0.9159 & 21.34 & 2.462 & 11.10 & 18.95 & 10.67$\pm$0.04 & 166$\pm$12 & 10.90$\pm$0.08 & 1.0 & 0.0\\
12019899 & 0.9325 & 21.71 & 0.469 & 11.34 & 15.63 & 10.80$\pm$0.06 & 236$\pm$17 & 10.48$\pm$0.08 & 1.6 & 0.2\\
12024453 & 0.9056 & 20.91 & 1.601 & 11.60 & 17.60 & 10.98$\pm$0.05 & 186$\pm$10 & 10.80$\pm$0.07 & 1.3 & 0.2\\
\enddata
\tablecomments{\label{table1}The first five columns are properties from T07, where the effective radii and stellar masses have uncertainties of 0.06 dex and 0.2 dex, respectively. $M^{R}_\star$ refers to the stellar mass derived with photometry from the RAINBOW database. SB$_e$ is the rest-frame $B$-band average surface brightness within the effective radius. Age and metallicity are luminosity-weighted by the linear combination of SSP models. }
\end{deluxetable*}

\subsection{Velocity Dispersions}
To measure $\sigma$'s we used the {\tt MOVEL} code within the {\reduceme} distribution \citep{cardiel99}. This program computes an optimal template by creating a linear combination of model stellar template spectra that yields the best fit to the galaxy spectrum. The linear combination coefficients of the templates weigh their contribution to the total stellar population and provide an estimate of the age and metallicity distribution of the galaxy. Our single stellar population (SSP) templates were obtained from the MILES library \footnote{miles.iac.es} \citep{sanchezblazquez06,vazdekis10} assuming Chabrier IMF and broadened to match the resolution of our galaxy spectra. These templates were selected with metallicities [Fe/H]$=\{0.0,0.2\}$ and a range of ages $0.5-5$ Gyr. We ruled out lower metallicities since they are very rare given the masses of our galaxies \citep{zahid11}. \\We fitted each spectrum using a rest-frame range of approximately $3700-4400$ \AA\, since the noise generated by the subtraction of sky OH lines washed out most features in the redder part of the observed wavelength range. Where applicable, we excluded the [OII] emission and uncorrected telluric absorption from the {\tt MOVEL} fits. In order to account for the dominant photon noise in the final $\sigma$ results, we performed Monte Carlo computations with 200 bootstrapped spectra. The output produced a distribution of values, whose mean and standard deviation give the value and error of the $\sigma$. For the objects 12019899 and 12024453 we calculated $\sigma$ as the weighted mean of the values derived from the two different instrumental resolutions, which favors the higher resolution value due to its smaller error. The fits to the spectra are displayed in Fig. \ref{fig1}. Following \citet{jorgensen95}, we apply an aperture correction to a diameter of 1.19 $\ensuremath{ h}^{-1}$ kpc, which increases our values around 6\%. The final $\sigma$'s are given in Table \ref{table1}. We note that there exists publicly available DEEP2 spectra for all of our galaxies \citep{davis07}. The target 12024453 has already been part of the photometric and spectroscopic study by \cite{fernandezlorenzo11}, who find $\sigma=173\pm16$ \kms from a spectrum of $\mathrm{(S/N)}=11$ per \AA. This is consistent with our measurement. The rest of the DEEP2 spectra of galaxies in common with our sample have (S/N)$\leq9$ per \AA\ and cannot be used for reliable $\sigma$ measurements.

\subsection{Stellar Populations}
The age and metallicity of each galaxy were determined from the values of the SSP model with the largest coefficient in the linear combination that best fitted the galaxy spectrum (see values in Table 1). These values show relatively low ages in the range of $0.8-1.6$ Gyr, with uncertainties estimated to be around 0.4 Gyr. In addition, we calculated photometric stellar masses using the data from the RAINBOW database\footnote{rainbowx.fis.ucm.es} \citep{barro11}, which includes photometry for our galaxies in $\sim$30 bands covering a range from 150 nm up to 70 $\mu$m. We fit these data to stellar population models by \citet{bc03} with Chabrier IMF, assuming an extinction law from \citet{calzetti00}. We constrain the metallicity in the same way than for the {\tt MOVEL} templates. The results show solar metallicity for all galaxies and ages in the range of 0.7-1.7 Gyr. Such outcome is very similar to the results from {\tt MOVEL}.\\
The galaxies 12024790 and 12024321 have [OII]$\lambda$3727 \AA\ emission, implying ongoing or recent star formation. We combine the [OII] fluxes from our spectra with the rest-frame \ensuremath{k}-corrected 24 $\mu$m fluxes from RAINBOW to derive star formation rates (SFR) using the conversion from \citet{kennicutt09} based on Kroupa IMF. The total SFR$_{\mathrm{[OII]+IR}}$ for the [OII] emission galaxies is 9.7 and 22.6 $M_\odot\,\mathrm{yr}^{-1}$, respectively.

\section{DISCUSSION}
Dynamical masses are calculated under the assumption of homology as $M_{\mathrm{dyn}}=\beta \Reff \sigma^2/G$ with $\beta=5$ (see Table 1), following studies of local ETGs \citep{cappellari06}. As noted in $\S$ 2.2, the galaxy 12024790 shows a double peak in the [OII] emission. If such feature is due to rotation of a gaseous disk, it yields a projected velocity of $V_r \mathrm{sin}(i)\simeq165$ \kms, which is comparable to its measured $\sigma=156$ \kms. However, it is unlikely that this caused an important bias in the dynamical mass estimates. At $\z=0$ it has been found that $\beta=5$ is robust against rotation up to $V_r/\sigma \approx 1$ \citep{cappellari06, cappellari07}, and there is no clear evidence that this would be different at $\z=1$ \citep{wel08b}. Therefore, we assume that any deviations from homology due to possible rotational support are within the errors of our dynamical masses. \\Our galaxies have values that are systematically smaller than the originally published stellar masses by an average factor of $\sim$6. This is much larger than the combined error of photometric stellar mass and radius (typically a factor of $\sim$3). Note that for the galaxy 12024321 these masses are consistent within uncertainties. The stellar population models we used with the RAINBOW data yield stellar masses in the range of log$\Mstar=10.65-10.98$, on average a factor of $\sim$1.8 larger than the dynamical masses. The masses from RAINBOW are derived with the same IMF and stellar population models as in T07 but with a broader spectral range (UV-FIR) as described in $\S$ 2.4. In order to investigate the large differences between both sets of masses, we also derived masses using only $BRIJK$ bands as in T07. We found no significant difference between our UV-FIR and $BRIJK$ masses, implying that the discrepancy with T07 cannot be simply explained by the use of a different spectral range. We also checked that our mass determinations are weakly sensitive to the choice of extinction model and small variations in metallicity. \citet{barro11b} derive masses for hundreds of galaxies in T07's sample using the full RAINBOW photometry. They find that, on average, their masses are consistent with those from T07. Therefore, the cause of the inconsistency with our photometric masses is unclear. Given the large uncertainties in the systematics of photometric masses, we adopt dynamical masses as the correct total mass estimates. Following local lensing studies \citep{gavazzi07}, we apply a fiducial dark matter fraction $f_{\mathrm{DM}}=0.3$ and recalculate the stellar masses as $0.7\times M_{\mathrm{dyn}}$. \\Comparison studies between stellar and dynamical mass of less dense ETGs than those in our sample have been conducted at low \citep{cappellari06} and high redshifts \citep{wel06} with the result of dynamical masses being on average equal or larger than stellar masses, as expected when considering a dark matter contribution. When \citet{wel06} calculate stellar masses using similar multi-band photometry and stellar population parameters as in T07, they find $M_\star/M_{\mathrm{dyn}}=0.66$ (Salpeter IMF). Therefore, our results point towards a systematic overestimation of the stellar masses used to select the galaxies of our sample. We cannot rule out a deviation from homology that requires $\beta \neq 5$. However, the theoretical expectation for objects with such high S\'ersic index is $\beta \lesssim 5$ \citep{cappellari06,bertin02}, which would yield even smaller dynamical masses. \\Fig. \ref{fig2}a shows the position of our objects in the stellar mass-size relation with the old/new stellar mass estimates. Fig. \ref{fig2}b shows our objects in the Fundamental Plane (FP). At a given $\Reff$, they have systematically larger masses and smaller combination of surface brightness (SB$_e$) and $\sigma$ than the respective local relations. Moreover, our objects follow a trend in which they lie further from both local relations at smaller radii. This result has also been found in $\z\sim0.9$ clusters by \citet{jorgensen06,jorgensen07}. Compared to the FP derived by these authors, at a given $\Reff$ our field sample shows a similar slope and a relative offset of $\sim0.3$ mag towards brighter $\mathrm{SB}_e$, which is in agreement with field galaxies having generally younger stellar populations than clusters of the same redshift.
\\In order to understand how our galaxies relate to low redshift ETGs, we investigate the simple evolutionary scenario based on passive luminosity fading of the stellar populations and growth via minor mergers that has been proposed for the general population of ETGs at $\z \sim 1$ \citep{gebhardt03, naab09}. To derive the amount of fading for our sample, we use \citet{bc03} models with Chabrier IMF and the redshift, ages and metallicities from Table \ref{table1}. Passive evolution from $\z \sim1$ to $\z=0$ results in an average dimming of $\sim$2 mag in SB$_e$. We assume that the passive luminosity evolution does not affect $\Reff$. Minor mergers are very efficient growing the size at a given mass increment, slightly decreasing $\sigma$. According to the simulations by \citet{naab09}, an ETG with $M_\star =10^{11}M_\odot$ and $\Reff=1.5$ kpc would grow a factor of 1.5 in mass via minor mergers from $\z \sim1$ to $\z=0$. We adopt this factor of mass growth for our galaxies. Following these authors, we model each merger as $\sigma^2_f/\sigma^2_i=(1+\eta \epsilon)/(1+\eta)$ and ${\Reff}_f/{\Reff}_i=(1+\eta)^{2}/(1+\eta \epsilon)$, where the subindexes (i)/(f) denote initial/final values of the host galaxy and $\epsilon=\sigma^2_a/\sigma^2_i$ and the mass ratio $\eta=M_a/M_i$ take into account the accreted system ($M_a,\sigma_a$). We assume a history of four mergers with constant $\eta=0.1$ and $\epsilon=0.2$ for all galaxies (note that $\eta$ and the total mass growth set the total number of mergers). This choice of values is consistent with the simulations by \citet{naab09}. The total change for $\Reff$ and $\sigma$ become factors of 2 and 0.8, respectively. These results are robust against variations in the values of $\eta$ and $\epsilon$ at the given total mass growth. We assume that the mass-to-light ratio does not change significantly after each merger. In addition, we assume that as the size increases, SB$_e$ decreases by $\Delta \mathrm{SB}_e\propto2.94\Delta \mathrm{log}\Reff$ \citep{hamabe87}. \\Fig. \ref{fig2} shows the combined evolution of minor mergers and luminosity fading for our galaxies. Most of them reach positions that are consistent with the local FP and mass-size relations, considering the estimated uncertainties in our measurements. Therefore, this simple evolutionary scenario can plausibly describe the evolution in our sample galaxies to become normal ETGs at $\z=0$. 

\acknowledgments This work is partially funded by the Spanish MICINN under the Consolider-Ingenio 2010 Program grant CSD2006-00070: First Science with the GTC. JMM acknowledges funding from a Graduate Alumni Fellowship at the University of Florida. MB and AH acknowledge support from the Spanish MICINN through grant AYA2009-11137. We thank the anonymous referee for useful comments that improved the presentation of this Letter. We are grateful to the GTC personnel who took the observations and to Ana Matkovic for her help with the use of the $\reduceme$ software.  \\

\end{document}